\documentclass[a4paper,12pt,reqno]{amsart}

\usepackage{amsmath,amsfonts,amsthm,amssymb}

\newcommand{\eps}{\varepsilon}

\makeatletter
\newcommand{\eqcolon}{\mathrel{\mathord{=}\raise.2\p@\hbox{:}}}
\newcommand{\coloneq}{\mathrel{\raise.2\p@\hbox{:}\mathord{=}}}
\makeatother

\newcommand{\RR}{\mathbb{R}}
\newcommand{\NN}{\mathbb{N}}

\newcommand{\CC}{\mathbb{C}}

\newcommand{\R}{\mathbb R}

\newtheorem{theorem}{Theorem}[section]

\newtheorem{corollary}[theorem]{Corollary}

\newtheorem{lemma}[theorem]{Lemma}

\newtheorem{remark}[theorem]{Remark}

\usepackage{pstricks} 
\usepackage{pst-node}
\usepackage{pst-tree}

\newcommand{\tnode}{\TR{\raisebox{0.5pt}{\ensuremath{\bullet}}}}

\psset{levelsep=-8pt,nodesep=-2pt,treesep=3pt}

\newcommand{\TT}{\mathcal{T}}
\newcommand{\VV}{\mathcal{V}}
\newcommand{\WW}{\mathcal{W}}
\newcommand{\BT}{\mathcal{BT}}
\newcommand{\troot}{\bullet}
\begin{document}

\title{Rooted trees for  3d Navier-Stokes equation}
\author{Massimiliano Gubinelli}
\address{Dip. di Matematica Applicata ``U. Dini'' -- Universit\`a di
Pisa, Italia}
\date{March 2006 (rev. 1)}
\maketitle

\begin{abstract}
We establish a representation of a class of solutions of 3d Navier-Stokes equations in $\R^3$ using sums over rooted trees.
We study the convergence properties of this series recovering in a simplified manner some results obtained recently by Sinai and other known results for solutions in spaces of pseudo-measures introduced initially by Le~Jan and Sznitman. The series representation make sense also in the critical case where there exists global solutions for small initial data and it allows the study of their long-time or small-distance behavior.
\end{abstract}

\section{Introduction}

We consider the NS equation in the form
\begin{equation}
  \label{eq:NS-basic}
v_t(k) = e^{-|k|^2 t} v_0(k) + i \int_0^t e^{-|k|^2 (t-s)} \int_{\RR^3}dk' \langle k, v_s(k-k')\rangle P_k v_s(k') \,ds
\end{equation}
where $v_t \in C(\RR^3 ; \mathbb{C}^3)$ is the Fourier transform of
the velocity field, $\langle \cdot ,\cdot \rangle$ is the scalar
product in $\mathbb{C}^3$ and $P_k: \CC^3 \to \CC^3$ is the projection
on the directions orthogonal to the vector $k \in \RR^3$, i.e.
$
P_k a = a - \langle k,a\rangle k |k|^{-1}
$.
Eq.~(\ref{eq:NS-basic}) will be studied in the spaces
$\Phi(\alpha,\alpha)$, $\alpha \in [2,3)$ and where $v \in \Phi(\alpha,\alpha)$ if $v \in C(\RR^3;\CC^3)$ with $k \cdot v(k) = 0$ and 
$$\|v\|_\alpha = \sup_{k\in\RR^3} |k|^\alpha |v(k)| < \infty. $$
We will write $\alpha = 2+\eps$ with $\eps \in [0,1)$.

The spaces $\Phi(\alpha,\alpha)$ are interesting because, in general,
they contain solutions with infinite energy and enstrophy so classical
results about existence and uniqueness of solutions
do not apply.

In a series of papers, Sinai~\cite{Sinai1,Sinai2,Sinai3}, studies
eq.~(\ref{eq:NS-basic}) in these spaces giving elementary proofs that in
$\Phi(\alpha,\alpha)$ with $\alpha > 2$, there is existence of unique local
solutions and that these solutions survive for arbitrary large time if
the initial condition is small enough.

Moreover in the ``critical'' space $\Phi(2,2)$ there is existence and
uniqueness of
global solutions for small initial data. This latter global result
was initially proven  by Le Jan and Sznitman~\cite{LS} using a probabilistic representation (under the name of \emph{stochastic cascades})
and afterwards reproved by Cannone and Planchon~\cite{CP} in a more standard
functional-analytic fashion. 

The analysis of the equation~(\ref{eq:NS-basic}) in various function spaces similar to $\Phi(\alpha,\alpha)$ is summarized in the work of  Bhattacharya et al. in~\cite{MR1997593}. A more recent review of the current status of the stochastic cascades approach has been recently given by Waymire in~\cite{MR2121794}.

We are interested in explicit series representations for these (local or global) solutions.
When $\alpha > 2$ Sinai~\cite{Sinai2} proved  that the local solution can be
represented by a absolutely convergent series and in~\cite{Sinai3} he
analyzed this series with the aim of understanding better the growth of
the various terms.
A different series representation  appears also in the book of Gallavotti~\cite{Gall}.

Our contribution will be to prove yet another series representation
for the solutions in $\Phi(\alpha,\alpha)$ including the critical case
$\alpha=2$ which was left open by the analysis of Sinai. This 
series representation is indexed by \emph{rooted trees}. 

Rooted trees appear naturally in the series expansion of solutions to
ODEs. They possess remarkable algebraic properties which were
masterfully exploited by Butcher~\cite{Butch} to provide a general
theory of Runge-Kutta (R-K) methods for numerical integration. 
Afterwards rooted trees appeared also in the work of Connes and Kreimer~\cite{CK}
on the mathematical structure of renormalization in quantum field
theory. The work of Brouder~\cite{Broud,Bit} gives  a short overview
of the algebraic properties of rooted trees and explore some connections
between R-K methods and renormalization.  

These widespread applications of rooted trees were the
initial motivations for this work. In the following we show that
rooted trees are a natural language in which the known results (and
some new ones) about
the eq.~(\ref{eq:NS-basic}) in the spaces $\Phi(\alpha,\alpha)$ can be
proven rather easily. Moreover the representation with a series
indexed by rooted-trees can be controlled in a straightforward way  and provide   informations on the solutions themselves, like the behavior for large times or for large wave-vectors.

\bigskip
The plan of the paper is as follows.
In Sec.~\ref{sec:trees} we introduce rooted trees which will be
used in Sec.~\ref{sec:series} to prove the series representation for the
solutions of NS equation. Next in Sec.~\ref{sec:remarks} we make some
observations on the different nature of some classes of terms which
contribute to the series and which appeared originally in the work of Sinai~\cite{Sinai3}. In App.~\ref{sec:remarks2} we review briefly, for sake
of completeness, the question of existence and uniqueness problem for
the equation~(\ref{eq:NS-basic}) in the spaces
$\Phi(\alpha,\omega)$. At the end, App.~\ref{sec:proofs} collects some proofs.

\section{Trees}
\label{sec:trees}
A rooted tree is a graph with a special vertex called \emph{root} such
that there is a unique path from the root to any other vertex of the
tree. Here some examples of rooted trees:
$$
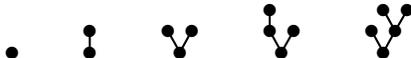
\tnode \qquad 
\pstree{\tnode}{\tnode}
\qquad
\pstree{\tnode}{\tnode \tnode}
\qquad
\pstree{\tnode}{\pstree{\tnode}{\tnode} \tnode}
\qquad
\pstree{\tnode}{\tnode \pstree{\tnode}{\tnode \tnode}} 
$$
We draw the root  at the bottom with the tree growing upwards (as real trees).
In a rooted tree the order of the branches at any vertex is ignored so
the following two are representations of the same tree:
$$
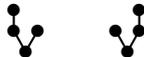
\pstree{\tnode}{\pstree{\tnode}{\tnode} \tnode}
\qquad
\pstree{\tnode}{\tnode \pstree{\tnode}{\tnode} }
$$
Given $k$ rooted trees $\tau_1,\cdots,\tau_k$ we define
$\tau = [\tau_1,\cdots,\tau_k]$ as the tree obtained by attaching the $k$
roots of $\tau_1,\cdots,\tau_k$ to a new vertex which will be the root
of $\tau$. Any tree can be constructed using the simple tree $\bullet$
and the operation $[\cdots]$, e.g.
$$
[\bullet] = \pstree{\tnode}{ \tnode}
\qquad
[\bullet,[\bullet]] = \pstree{\tnode}{\pstree{\tnode}{\tnode} \tnode},
\qquad \text{etc\dots}
$$
On trees we will define various functions. Denote $\TT$ the set of all
rooted trees and let $|\cdot| : \TT \to \NN$ the map which counts the
number of vertices of the tree and which can be defined
recursively as
$$
|\bullet | = 1, \qquad |[\tau_1,\dots,\tau_k]| = 1+|\tau_1|+\cdots+|\tau_k|
$$
moreover we define the \emph{tree factorial} $\gamma: \TT \to \NN$ as
$$
\gamma(\bullet) = 1, \qquad \gamma([\tau_1,\dots,\tau_k]) =
|[\tau_1,\dots,\tau_k]| \gamma(\tau_1) \cdots \gamma(\tau_k).
$$
Last, we define the symmetry factor $\sigma : \TT \to \NN$: this is
defined recursively as $\sigma(\bullet) = 1$ and $\sigma([\tau_1
\cdots \tau_n]) = s(\tau_1,\dots,\tau_n) \sigma(\tau_1)\cdots
\sigma(\tau_n)$ where $s(\tau_1,\dots,\tau_n)$ is the order of the
permutation group of the (ordered) $n$-uple $(\tau_1,\dots,\tau_n) \in
\TT^n$. In the sequel we will only need to consider the subset
$\BT \subset \TT$ which contains rooted trees with at most
two sons for each vertex, this is due to the bilinear nature of the
non-linear term in the NS equation. 

\section{Series representation}
\label{sec:series}
If we let $c_t(k) = |k|^\alpha v_t(k)$ the eq.~(\ref{eq:NS-basic}) above takes the form
\begin{equation}
  \label{eq:NS-c}
c_t(k) = e^{-|k|^2 t} c_0(k) + i \int_0^t e^{-|k|^2 (t-s)} |k|^\alpha
\int_{\RR^3}dk' \frac{\langle k, c_s(k-k')\rangle P_k
c_s(k')}{|k-k'|^{\alpha} |k'|^{\alpha}} 
\end{equation}
for function $c \in C(\RR^3;\CC^3)$ such that $\sup_t \|c_t\|_0 < \infty$, $\langle k,c(k) \rangle = 0$ and $c(0) = 0$.

For simplicity write the above equation in the abstract form
\begin{equation}
  \label{eq:NS-c-abstract}
c_t = S_t c_0 +  \int_0^t S_{t-s}  B(c_s,c_s)\,ds.
\end{equation}
where $S$ is a bounded semigroup and $B$ is the symmetrized bilinear operator
 $ B(c,d) = (B_1(c,d) + B_1(d,c))/2$ with
$$
B_1(c,d) = i  |k|^\alpha
\int_{\RR^3}dk' \frac{\langle k, c(k-k')\rangle P_k
d(k')}{|k-k'|^{\alpha} |k'|^{\alpha}}. 
$$

Let $\VV = \{ c \in C(\RR^3;\CC^3) : c(0) = 0, \langle k,  c(k) \rangle = 0\; \text{and}\; \|c\|_0 <
\infty \}$ and for any $T > 0$ define the Banach space $\WW_T = C_b([0,T], \VV)$ endowed
with the sup norm.  
Define the bilinear operator $\mathcal{B} : \WW_T\otimes \WW_T \to \WW_T$ as
\begin{equation*}
\mathcal{B}_t(c,d) = \int_0^t S_{t-s}  B(c_s,d_s)\,ds.  
\end{equation*}

\begin{lemma}
For any $\alpha \ge 2$ and any $T > 0$, the operator $\mathcal{B}$ is well defined and there exists an
increasing function
$N : \RR_+ \to \RR_+$ such that
\begin{equation}
  \label{eq:estimate-B}
|\mathcal{B}_t(c,d)| \le N_t \|c\|_0 \|d\|_0  
\end{equation}
where  $N_t$ tends to zero as $t \to 0$ for any $\alpha
\ge 2$. Moreover when $\alpha = 2$ we have a uniform bound $\sup_{t \ge
  0} N_t \le  N_* <\infty $ independent of $T$.
\end{lemma}
\begin{proof}
The proof can be found in the paper of Sinai~\cite{Sinai1} and 
consists in a direct estimate of the integral. Some
general considerations on the bilinear operator are summarized in App.~\ref{sec:remarks2}.  
\end{proof}

Now define the operator $\phi : \BT \times \VV \to \WW_T$ by recurrence
as
\begin{equation}
  \label{eq:3}
\phi(\troot;h) = \mathcal{B}(S_{\cdot} h,S_{\cdot }h)
\qquad
\phi([\tau];h) = 2 \mathcal{B}(S_{\cdot} h,\phi(\tau;h))
\end{equation}
and
\begin{equation}
  \label{eq:3c}
\phi([\tau_1 \tau_2];h) = 2 \mathcal{B}(\phi(\tau_1;h),\phi(\tau_2;h))
\end{equation}
for any $h\in \VV$,$\tau,\tau_1,\tau_2 \in \BT$ where we let $(S_{\cdot} h)_t = S_t h$.
 
To find solutions of eq.~(\ref{eq:NS-c-abstract}) in $\WW_T$ with initial
condition $h \in \VV$ we set up Picard iterations $\{u^{(n)} \in \WW_T \}_n$ as $u^{(0)}_t =
S_t h$ and
$
u^{(n+1)}_t = S_t h + \mathcal{B}_t(u^{(n)},u^{(n)})
$.

\begin{lemma}
The functions $u^{(n)}$ have the representation
\begin{equation}
  \label{eq:rep-finite}
u^{(n)} = S_{\cdot} h + \sum_{\tau \in \BT_{n-1}} \frac{1}{\sigma(\tau)} \phi(\tau;h)  
\end{equation}
where $\BT_n \subset \BT$ is the set of rooted trees for which the leaves
are at distance at most $n$ from the root and where we conventionally let $\BT_{-1} =
\emptyset$. 
\end{lemma}
\begin{proof}
It is clear that the formula holds for $n = 0$ (the sum does not
contain any terms). Assume it holds for any $k \le n$ and let us prove
it for $n+1$:
\begin{equation*}
  \begin{split}
u^{(n+1)} & = S_\cdot h +  \mathcal{B}(u^{(n)},u^{(n)})
  \\ & = S_\cdot h + \mathcal{B}(S_\cdot h, S_\cdot h) +  2
  \sum_{\tau \in \BT_{n-1}} \frac{1}{\sigma(\tau)} \mathcal{B}(S_\cdot h,\phi(\tau;h)) 
\\ & \qquad 
+
  \sum_{\tau^1,\tau^2 \in \BT_{n-1}}
  \frac{1}{\sigma(\tau^1)\sigma(\tau^2)} \mathcal{B}(\phi(\tau^1;h),\phi(\tau^2;h))
  \\ & = S_\cdot h +  \phi(\troot;h)
  +  
  \sum_{\tau \in \BT_{n-1}} \frac{1}{\sigma(\tau)} \phi([\tau];h)
\\ & \qquad
 +
  \sum_{\tau^1,\tau^2 \in \BT_{n-1}, \tau^1 \neq \tau^2} 
  \frac{1}{2\sigma(\tau^1)\sigma(\tau^2)} \phi([\tau^1 \tau^2];h)
+   \sum_{\tau \in \BT_{n-1}}
 \frac{1}{2 \sigma(\tau) \sigma(\tau)}
  \phi([\tau \tau];h)
  \\ & = S_\cdot h + \sum_{\tau \in \BT_{n}} \frac{1}{\sigma(\tau)} \phi(\tau;h)
  \end{split}
\end{equation*}
since $\sigma([\tau \tau]) = 2 \sigma(\tau)^2$ and $\sigma([\tau]) =
\sigma(\tau)$.    
\end{proof}

The norm convergence of the series
\begin{equation}
  \label{eq:main-series}
u =  S_{\cdot} h + \sum_{\tau \in \BT} \frac{1}{\sigma(\tau)} \phi(\tau;h)    
\end{equation}
in $\WW_T$ implies  convergence of the Picard iterates $u^{(n)}$ to the
element $u \in \WW_T$ which satisfy eq.~(\ref{eq:NS-c-abstract}) in
$[0,T]$ with
initial condition $u_0 = h$.

\bigskip

Define the following function $\theta : \BT \to \RR$:
\begin{equation*}
\theta(\troot) = 2, \qquad \theta([\tau]) = 1+ \theta(\tau), \qquad
\theta([\tau_1 \tau_2]) = \theta(\tau^1) + \theta(\tau^2)  
\end{equation*}
and note that $h \mapsto \phi_{ts}(\tau; h)$ is an homogeneous
function of order $\theta(\tau)$. Always holds
\begin{equation}
  \label{eq:theta-bound}
 (|\tau|+1)/2 \le
\theta(\tau) \le
|\tau|+1  
\end{equation}
 as easily proven by induction on $|\tau|$.

\bigskip
The following control of the coefficients of the series~(\ref{eq:main-series}) is the main result of this note.

\begin{theorem}
\label{th:main}
For any $\eps \in [0,1)$ the following estimate is true
\begin{equation*}
|\phi_{t}(\tau;h)(k)| \le C_{\tau} e^{-|k|^2 t / (|\tau|+1)}
t^{|\tau| \eps /2} \|h\|_0^{\theta(\tau)}  
\end{equation*}
where the constants $C_{\tau}$ satisfy:
$$
C_{[\tau^1 \tau^2]} = \frac{A}{|[\tau^1 \tau^2]|^{\eps/2}} C_{\tau^1} C_{\tau^2}, \qquad C_{[\tau]} = \frac{A}{|[\tau]|^{\eps/2}}
C_{\tau}, \qquad C_{\troot} = A
$$
for some constant $A$ depending only on $\alpha$. 
\end{theorem}
The proof of this theorem is reported in App.~\ref{sec:proofs}.

\begin{remark}
\label{rem:chooseC}
The constants $C_\tau$ can be chosen as follows:
$$
C_\tau = A^{|\tau|} \gamma(\tau)^{-\eps/2}.
$$  
\end{remark}

Now we can prove the following result about existence and
series representation of solutions of eq.~(\ref{eq:NS-basic}) in the
spaces $\Phi(\alpha,\alpha)$ ($\alpha=2$ included).
\begin{corollary}
\label{cor:1}
The series~(\ref{eq:main-series}) has the following properties:
\begin{itemize}
\item[a)] for $\eps \in(0,1)$ and fixed $\|h\|_0$ it converges in norm for small
$t_*$ and solve the problem~(\ref{eq:NS-c}) in $\WW_{t_*}$;
\item[b)] for $\eps \in(0,1)$  and fixed $T$ it converges in norm in $\WW_{T}$ for $\|h\|_0$ small
enough;
\item[c)] for $\eps = 0$ (i.e. $\alpha =2$) and for $\|h\|_0$ small enough it converges in norm in $\WW_\infty$ and define a global solution of the
problem~(\ref{eq:NS-c}).
\end{itemize}
\end{corollary}
\begin{proof}
Using Thm.~\ref{th:main} and Remark~\ref{rem:chooseC} we find that there exists a constant $B$ such that
\begin{equation}
\label{eq:first-bound}
  \begin{split}
|u_t(k) - [S_t h](k)|& \le \sum_{\tau} B^{|\tau|} \sigma(\tau)^{-1} \gamma(\tau)^{-\eps/2}  
  e^{-|k|^2 t / (|\tau|+1)}
t^{|\tau| \eps /2} \|h\|_0^{\theta(\tau)} 
\\ & \le \sum_{n \ge 1} Z_n B^n
e^{-|k|^2 t/(n+1)} t^{n \eps/2} \|h\|_0^{(n+1)/2} \left(1 \wedge \|h\|_0^{(n+1)/2} \right).      
  \end{split}
\end{equation}
where $Z_n$ is the number of rooted trees in $\BT$ with $n$ vertices.
The following recursive relations can be used to bound the $Z_n$'s:
$$
Z_1 = 1, \qquad Z_{n+1} \le Z_n +  \sum_{n_1+n_2 = n} Z_{n_1} Z_{n_2}.
$$
From this relation it is not difficult to prove that $Z_n$ grows at most exponentially, i.e. there exists a constant $D$ such that
\begin{equation}
  \label{eq:Z-bound}
Z_n \le D^n (n+1)^{-3/2}.  
\end{equation}
(see for example~\cite{Sinai2}, Sec.3).

Next, by induction we can prove that $\gamma(\tau) \ge 2^{|\tau|-1}$. This bound is optimal since it is saturated by the binary trees for which every path from the root to the leaves has the same length.
Using this bound, eq.~(\ref{eq:theta-bound}) and eq.~(\ref{eq:Z-bound}) in eq.~(\ref{eq:first-bound}) we get
\begin{equation}
\label{eq:first-bound-2}
  \begin{split}
|u_t(k)| & \le  |[S_t h](k)|+  \sum_{n \ge 1} Z_n B^n
e^{-|k|^2 t/(n+1)} t^{n \eps/2} \|h\|_0^{(n+1)/2} \left(1 \wedge \|h\|_0^{(n+1)/2} \right)
\\ & \le  \|h\|_0 +  \sum_{n \ge 1}  (DB t^{\eps/2})^n (n+1)^{-3/2}
 \|h\|_0^{(n+1)/2} \left(1 \wedge \|h\|_0^{(n+1)/2} \right).       
  \end{split}
\end{equation}
so the series~(\ref{eq:main-series}) converges in norm whenever the geometric series
$$
 \sum_{n \ge 1}  (DB t^{\eps/2})^n 
 \|h\|_0^{(n+1)/2} \left(1 \wedge \|h\|_0^{(n+1)/2} \right)
$$
converges. This gives directly  $a),b),c)$. Indeed note that for $\eps = 0$ the dependence on $t$ disappear in this last series.
\end{proof}

Now, lets come back to the original variables. It is clear that the function $v_t(k) = |k|^{-2} u_t(k)$ satisfy eq.~(\ref{eq:NS-basic}) in $[0,T]$ when the series defining $u$ converges in $\WW_T$.  Here we are interested in the behavior of the global solutions when $\alpha =2$:

\begin{corollary}
In the case $\alpha = 2$ and when $\|h\|_2$ is sufficiently small the
global solution $v$ of  eq.~(\ref{eq:NS-basic}) with initial condition
$h$ has the following two properties:
\begin{itemize}
\item[a)]  for fixed $k \in \RR^3\backslash \{0\}$, $
\lim_{t \to \infty } |v_t(k)| = 0  $;
\item[b)] for fixed $t > 0$, there exists two constants $C_3,C_4$ such that
 $|v_t(k)| \le C_3 e^{-C_4 |k|\sqrt{t}}$ as $|k| \to \infty$.
\end{itemize}
\end{corollary}

\begin{proof}
By the same bounds performed in Cor.~\ref{cor:1} we see that the function $v_t(k) = |k|^{-2} u_t(k)$ satisfy the inequality 
\begin{equation*}
|v_t(k)| \le  e^{-|k|^2 t} |h(k)| + \sum_{n \ge 1} C_1^n 
(n+1)^{-3/2} e^{-|k|^2 t/(n+1)} \|h\|_2^{(n+1)/2}   , \qquad k \neq 0
\end{equation*}
for $\|h\|_2$ small enough to guarantee the convergence of the series
$$
\sum_{n \ge 1} C_1^n 
(n+1)^{-3/2} \|h\|_2^{(n+1)/2}.
$$
Then fixed $k \in \RR^3\backslash\{0\}$ we have   $\lim_{t \to \infty} e^{-|k|^2 t/(n+1)} = 0$ for each $n$ and we obtain that $|v_t(k)| \to 0$ as $t \to \infty$.

Next, we want to estimate the series at fixed $t$ and for $|k| \to \infty$ by Laplace method.
Write
$$
\sum_{n \ge 1} C_1^n 
(n+1)^{-3/2} e^{-|k|^2 t/(n+1)} \|h\|_2^{(n+1)/2} \le \|h\|_2^{1/2}
\sum_{n \ge 1}  e^{-|k|^2 t/(n+1) + n \log(C_1 \|h\|_2^{1/2})}  
$$
The exponent in the sum of the r.h.s has a maximum for $n \simeq |k|\sqrt{t}/\sqrt{|\log(C_1 \|h\|_2^{1/2})|}$ and so, when $|k| \to \infty$ we have
$$
\sum_{n \ge 1}  e^{-|k|^2 t/(n+1) + n \log(C_1 \|h\|_2^{1/2})}  \le C_3 e^{-C_4 |k|\sqrt{t}}
$$
where $C_4 = 2/\sqrt{|\log(C_1 \|h\|_2^{1/2})|}$ and $C_3$ is some finite constant.
\end{proof}

\subsection{Remarks on particular subsets of trees}
\label{sec:remarks}
In the series~(\ref{eq:main-series}) different classes of trees give different
contributions. We define \emph{simple} trees the trees with at most
one branch at each vertex, i.e. of the form $[\cdots [\bullet]\cdots
]$. \emph{Short} trees are instead trees for which at  each  vertex we
have two branches, each of which carries a fixed proportion ($\alpha$ or $1-\alpha$) of the
vertices. Of course this will not be possible in general, so we allow
the proportion to oscillate around $\alpha$ in the interval
$[\alpha-\Delta \alpha, \alpha + \Delta \alpha]$, for some fixed $0 <
\Delta \alpha < \min(\alpha,1-\alpha)$. Since rooted trees does not
distinguish between branches at a vertex, we take here the convention
that the branches are ordered by the number of vertices in the
corresponding subtree. With this convention we can consider, without
loosing generality, values of $\alpha \in (0,1/2)$.

We will denote $\BT_0$ the set of simple trees and $\BT_\alpha$ the
set  of short trees corresponding to the proportion $\alpha$. 

  We have a
first simple lemma:
\begin{lemma}
\label{lemma:comb}
For $\tau \in \BT_0$ we have $\gamma(\tau) = |\tau|!$. For any $\alpha
\in (0,1/2)$ there exists constants $D_1,D_2,D_3,D_4$ such that, for any
$\tau \in \BT_\alpha$ we have
$$
 D_3   |\tau|^{-1} D_4^{|\tau|} \le \gamma(\tau) \le D_1   |\tau|^{-1} D_2^{|\tau|}.
$$  
\end{lemma}
\begin{proof}
The proof of the first claim is trivial. For the second, note that we
can choose the constants $D_1,D_2,D_3,D_4$ such that the inequalities are true for all
the trees $\tau \in \BT_\alpha$  with $|\tau| \le \overline n$ for some fixed
$\overline n$ and moreover they satisfy
\begin{equation*}
\frac{D_1 D_2^{-1}}{(\alpha-\Delta\alpha)(1-\alpha-\Delta\alpha-\overline n^{-1})}  \le 1
\quad\text{and}\quad
\frac{D_3 D_4^{-1}}{(1-\alpha+\Delta\alpha)(\alpha+\Delta\alpha)}  \ge 1.
\end{equation*}

 Then we proceed by induction on $n \ge \overline n$. Assume the inequality is true for trees with $|\tau| < n$ and  observe that, for $\tau \in
\BT_\alpha$, $|\tau| = n$ we have $\tau = [\tau_1 \tau_2]$ with 
$$
(\alpha-\Delta\alpha) \le \frac{|\tau_1|}{|\tau|} \le (\alpha+\Delta\alpha)
$$
and
$$
(1-\alpha-\Delta\alpha-|\tau|^{-1}) \le \frac{|\tau_2|}{|\tau|} \le (1-\alpha+\Delta\alpha-|\tau|^{-1})
$$
Then
\begin{equation*}
  \begin{split}
\gamma(\tau) & = |\tau| \gamma(\tau_1) \gamma(\tau_2)    
\le D_1^2 \frac{D_2^{|\tau_1|+|\tau_2|}}{|\tau_1||\tau_2|}
\le \frac{D_1^2 D_2^{-1}}{(\alpha-\Delta\alpha)(1-\alpha-\Delta\alpha-|\tau|^{-1})} \frac{D_2^{|\tau|}}{|\tau|} 
\\ & \le \frac{D_1^2 D_2^{-1}}{(\alpha-\Delta\alpha)(1-\alpha-\Delta\alpha-\overline n^{-1})} \frac{D_2^{|\tau|}}{|\tau|} 
 \le  D_1 \frac{D_2^{|\tau|}}{|\tau|} 
  \end{split}
\end{equation*}
and similarly we obtain
$
\gamma(\tau) \ge D_3 D_4^{|\tau|} |\tau|^{-1}
$,
proving the claim.
\end{proof}

This different behavior of the two classes of trees is responsible for
different convergence properties of the sum~(\ref{eq:main-series}) when
restricted to simple or short trees.

Define
$$
w_t = \sum_{\tau \in \BT_0} \frac{1}{\sigma(\tau)} \phi_{t}(\tau; h)
$$
then as consequence of Lemma~\ref{lemma:comb} and Theorem~\ref{th:main} we have the following result: 

\begin{corollary}
\label{cor:exp-decay}
For $\eps > 0$,
the series $w_t$ converges in $\VV$ for every $t$ and every initial condition
$h\in\VV$ and
$$
|w_t(k)| \le B' \sum_{n=1}^\infty  \frac{B^n}{n^{3/2} (n!)^{\eps/2}} e^{-|k|^2 t / (n+1)}
t^{n \eps /2} (1+\|h\|_0)^{n+1}. 
$$
\end{corollary}
\begin{proof}
The estimates on the series are similar to those in Cor.~\ref{cor:1}, but now the coefficient $\gamma(\tau) = |\tau|!$ goes to infinity fast enough to guarantee the convergence of the series for any time.  
\end{proof}

In~\cite{Sinai3}, Sinai studied different classes of contributions to
his series representation of NS. He calls the various contributions
\emph{diagrams} and then introduces short and simple diagrams which
are analogous to short and simple trees (even if diagrams does not
exactly corresponds to our trees). He then shows that the contribution
of the simple diagrams cannot cause the divergence of the overall
series. Corollary~\ref{cor:exp-decay} is the analogous of this result
in our setting.

For short trees the function $\gamma$ behaves exponentially with the
size of the tree and this is not enough to make the series restricted
to short trees converge for arbitrary time (when $\eps > 0$). 
A similar phenomenon is observed in~\cite{Sinai3} for short
diagrams.

\section*{acknowledgment}
The author wish to thanks the organizers and the lecturers of the CIME 2005 Summer
school ``SPDE in hydrodynamics: recent progress and prospects'' for the nice environment
and in particular Y.~Sinai whose lectures on the mathematics of
the Navier-Stokes system introduced the author to the problem studied in this note. Moreover useful comments by
Y. Bakhtin and by an anonymous referee are gratefully acknowledged.


\appendix
\section{Remarks on the spaces $\Phi(\alpha,\omega)$}
\label{sec:remarks2}
Without going in detailed proofs we would like to note some remarks
about the natural functional spaces in which solutions of
eq.~(\ref{eq:NS-basic})  live. Following Sinai~\cite{Sinai1} define the
space $\Phi(\alpha,\omega)$ ($\alpha,\omega \ge 0$) as the space of continuous functions $v :
\RR^3 \to \CC^3$ such that 
$$
\|v\|_{\alpha,\omega} = \sup_{k \in \RR^3} \psi(k)^{-1} |v(k)| <\infty
$$
where  $\psi(k) = |k|^{-\alpha}$ for $|k| \le 1$, $\psi(k) =
|k|^{-\omega}$ for $|k| \ge 1$.
Functions in this space can be bounded above by $|k|^{-\alpha}$ for
small $k$ and by $|k|^{-\omega}$ for large $k$. 
Consider the the bilinear integral operator
\begin{equation}
  \label{eq:bilin}
\widetilde{\mathcal{B}}(v,v)(t,k) = i \int_0^t e^{-|k|^2 (t-s)}  
 \int_{\RR^3}dk' \langle k, v_s(k-k')\rangle P_k v_s(k')\, ds 
\end{equation}
appearing in the r.h.s of eq.~(\ref{eq:NS-basic}). For this operator
we have the bound
\begin{equation}
  \label{eq:B-bound}
|\widetilde{\mathcal{B}}(v,v)(t,k)| \le \sup_{0 \le s \le t}
\|v_s\|_{\alpha,\omega}^2 |k|^{-1} (1-e^{-|k|^2 t}) I(k)  
\end{equation}
where
$
I(k) = \int_{\RR^3} dk' \psi(k-k') \psi(k)
$.
 For $k \neq 0$ the integral $I(k)$ converges when $\omega > 3/2$ and $\alpha < 3$ and we prove easily that, when $|k|
\le 1$
\begin{equation*}
I(k) \le C_\alpha \begin{cases}
1 & \text{for $\alpha < 3/2$}\\
|\log|k|| & \text{for $\alpha = 3/2$} \\
|k|^{3-2\alpha} & \text{for $3/2 < \alpha < 3$}
\end{cases}  
\end{equation*}
while when $|k|> 1$
\begin{equation*}
I(k) \le C_\omega \begin{cases}
|k|^{3-2\omega} & \text{for $3/2 <\omega < 3$}\\
|k|^{-\omega} & \text{for $\omega \ge 3$}
\end{cases}  
\end{equation*}
This behavior translates in the following estimates for the r.h.s. of eq.~(\ref{eq:B-bound}).
So when $|k|
\le 1$:
\begin{equation*}
|k|^{-1} (1-e^{-|k|^2 t}) I(k) \le C_\alpha \begin{cases}
 |k|^{-\alpha} t^{(1-\alpha)/2} & \text{for $\alpha < 1$}\\
 |k|^{-\alpha} (1 \wedge |t|) & \text{for $1 \le \alpha \le 2$}\\
 |k|^{-\alpha} |t|^{(\alpha-2)/2} & \text{for $2 \le \alpha < 3$}\\
\end{cases}  
\end{equation*}
and when $|k| \ge 1$
\begin{equation*}
|k|^{-1} (1-e^{-|k|^2 t}) I(k) \le C_\omega \begin{cases}
 |k|^{2-2\omega} (1-e^{-|k|^2 t}) & \text{for $3/2 < \omega < 2$}\\
 |k|^{-\omega} (1 \wedge |t|)^{(\omega-2)/2} & \text{for $2 \le \omega \le 3$}\\
 |k|^{-\omega} (1 \wedge |t|)^{1/2} & \text{for $3 \le \omega$}\\
\end{cases}  
\end{equation*}
These bounds imply that $\widetilde{\mathcal{B}}$ maps
$C([0,T],\Phi(\alpha,\omega))$ in itself whenever $\omega \ge 2$ for
any $\alpha \in [0,3)$ and in this case the norm $N$ of $\widetilde{\mathcal{B}}$ is given
by $$N_T = \sup_{t \le T} \sup_k [\psi(k)^{-1} |k|^{-1} (1-e^{-|k|^2
t}) I(k)]$$ and become  small
with $T$ allowing a direct proof of existence and uniqueness of
solutions to eq.~(\ref{eq:NS-basic}) for small time. Moreover when
$\alpha \in [1,2]$ the norm $N_T$ is
uniformly bounded in $T$ and this implies existence and uniqueness of
global solutions with small enough initial condition. 

Note moreover that the same bounds are true on the torus (only wave-vectors $|k| \ge 1$ are important in this case) and that they always imply
uniform control in time of the norm $N_T$ for any $\omega \ge 2$. 
In this case we have the existence and uniqueness of global solutions with
small initial conditions whose decay at infinity is not worse than $|k|^{-\omega}$.
Details can be found in~\cite{Sinai1}.

\section{Proofs}
\label{sec:proofs}
\subsection{Theorem~\ref{th:main}}
\begin{proof}
We will prove the statement by induction on $|\tau|= n$. Let us
assume that the estimate is true for any tree $\tau'$ with $|\tau'| <
n$ and let us prove it for trees $\tau$ with $|\tau| = n$. 
Consider the case $\tau = [\tau^1 \tau^2]$ with $|\tau^1| = p$,
$|\tau^2| = q$:
\begin{equation*}
  \begin{split}
\left|\phi_{t}([\tau_1 \tau_2];h)(k)\right|
&  \le 2 \int_0^t e^{-|k|^2 (t-u)} 
|B(\phi_{u}(\tau_1;h), \phi_{u}(\tau_2;h))|\,du    
\\
&  \le 2 \int_0^t du\,e^{-|k|^2 (t-u)} 
 |k|^{\alpha+1}
\int_{\RR^3}dk'\, \frac{|\phi_{u}(\tau_1;h)(k-k')|\, |\phi_{u}(\tau_2;h)(k')|}{|k-k'|^{\alpha} |k'|^{\alpha}}    
\\
&  \le 2 C_{\tau^1} C_{\tau^2} \|h\|_0^{\theta(\tau)}
\int_s^t du\, e^{-|k|^2 (t-u)} u^{\eps/2 (|\tau^1|+|\tau^2|)}
 |k|^{\alpha+1}
\\ & \qquad \cdot
\int_{\RR^3}dk' \frac{e^{-|k-k'|^2 u / (p+1)-|k'|^2 u / (q+1)}}{|k-k'|^{\alpha} |k'|^{\alpha}}.    
  \end{split}
\end{equation*}
The exponent in the integral has a maximum as a function of $k'$ and
$$
\frac{|k-k'|^2 u}{p+1} + \frac{|k'|^2 u}{q+1}  \ge \frac{|k|^2 u}{p+q+2} 
$$
for any $k' \in \RR^3$. So
\begin{equation}
\label{eq:this}
  \begin{split}
\left|\phi_{t}([\tau_1 \tau_2];h)(k)\right|
&   \le 2 C_{\tau^1} C_{\tau^2} \|h\|_0^{\theta(\tau)}
\\ & \qquad \cdot
e^{-|k|^2 t/(p+q+2)} \int_0^t du\, e^{-|k|^2 (t-u) (p+q+1)/(p+q+2)}  
 (u)^{\eps/2 (|\tau^1|+|\tau^2|)} |k|^{\alpha+1}
\\ & \qquad \cdot
\int_{\RR^3}dk' \frac{1}{|k-k'|^{\alpha} |k'|^{\alpha}}   
\\ &   \le A' C_{\tau^1} C_{\tau^2} \|h\|_0^{\theta(\tau)}
e^{-|k|^2 t/(p+q+2)}
\\ & \qquad \cdot
 \int_0^t du\, e^{-|k|^2 (t-u) (p+q+1)/(p+q+2)}  
(u)^{\eps/2 (|\tau^1|+|\tau^2|)}
 |k|^{4-\alpha}
  \end{split}
\end{equation}
where, if $e$ is a unit vector in $\RR^3$ we let
$$
A' = 2\int_{\RR^3}dk' \frac{1}{|e-k'|^{\alpha} |k'|^{\alpha}} < \infty.   
$$
Consider the term on this last line of eq.~(\ref{eq:this}):
\begin{equation*}
  \begin{split}
 \int_0^t  & du\, e^{-|k|^2 (t-u) (p+q+1)/(p+q+2)}   
(u)^{\eps/2 (|\tau^1|+|\tau^2|)}
 |k|^{4-\alpha}
 \\ & = t^{(|\tau^1|+|\tau^2|+1)\eps/2}
          \int_0^1 |\tilde k|^{4-\alpha} e^{-|\tilde k|^2 (1-u) (p+q+1)/(p+q+2)}
u^{\eps/2 (|\tau^1|+|\tau^2|)}\, du
  \end{split}
\end{equation*}
with $\tilde k = (t-s)^{1/2} k$. Let $a = (p+q+1)/(p+q+2)$.
 We
have the following bound, proved in lemma~\ref{lemma:bound-int} below:
\begin{equation*}
\int_0^1 e^{-|\tilde k|^2 (1-u) a} u^{(p+q)\eps/2} du \le 
\left(a |\tilde k|^2   + (p+q)\eps/2\right)^{-1}
\end{equation*}
Gathering all together we get
\begin{equation*}
  \begin{split}
\left|\phi_{t}([\tau_1 \tau_2];h)(k)\right|
&    \le A' C_{\tau^1} C_{\tau^2} \|h\|_0^{\theta(\tau)}
e^{-|\tilde k|^2/(|\tau|+1)} t^{|\tau|\eps/2} \frac{|\tilde k|^{2-\eps}}{a |\tilde k|^2  + (p+q)\eps/2}
  \end{split}
\end{equation*}
When $\eps \in (0,1)$ we have
$$
\sup_{\tilde k \in \RR^3} \frac{|\tilde k|^{2-\eps}}{a |\tilde k|^2  + (p+q)\eps/2} \le K [(p+q)]^{-\eps/2}
$$
where $K$ is a constant not depending on $\eps$, so in this case
\begin{equation*}
  \begin{split}
\left|\phi_{t}([\tau_1 \tau_2];h)(k)\right|
&    \le A'' C_{\tau^1} C_{\tau^2} \|h\|_0^{\theta(\tau)}
e^{-|\tilde k|^2/(|\tau|+1)} \frac{t^{|\tau|\eps/2}}{|\tau|^{\eps/2}}
  \end{split}
\end{equation*}
When $\eps = 0$ we have instead the bound
\begin{equation*}
  \begin{split}
\left|\phi_{t}([\tau_1 \tau_2];h)(k)\right|
&    \le A'' C_{\tau^1} C_{\tau^2} \|h\|_0^{\theta(\tau)}
e^{-|\tilde k|^2/(|\tau|+1)}
  \end{split}
\end{equation*}
Proving the claim in this case. The other cases can be treated similarly.
\end{proof}

\begin{lemma}
\label{lemma:bound-int}
$$
\int_0^1 e^{-a(1- u)} u^{b} du \le 1\wedge (a+b)^{-1}
$$  
\end{lemma}
\begin{proof}
Easy:
\begin{equation*}
  \begin{split}
\int_0^1 e^{-a (1-u)} u^{b} du & =
    \int_0^1 e^{-a u} (1-u)^{b} du
    \int_0^1 e^{-a u - b u}  du
\\  & \le
    \int_0^\infty e^{-(a+b)u } du
= (a+b)^{-1}
  \end{split}
\end{equation*}
\end{proof}

\end{document}